\documentclass[12pt,a4paper]{article}
\usepackage[utf8]{inputenc}
\usepackage[T1]{fontenc}
\usepackage[french]{babel}
\usepackage{amsmath, amssymb, amsfonts}
\usepackage{amsthm}
\usepackage{graphicx}
\usepackage{geometry}
\usepackage{multirow}
\usepackage{caption}

\theoremstyle{plain}
\newtheorem{theorem}{Theorem}[section]
\newtheorem{corollary}{Corollary}[section]
\newtheorem{definition}{Definition}[section]

\theoremstyle{remark}
\newtheorem{remark}{Remark}[section]
\title{Nonparametric Estimation of Extropy, Rényi Extropy, and Tsallis Extropy: \\
Almost Sure Convergence and Asymptotic Normality}
\author{Amadou Diadie BA\\
LERSTAD, Gaston Berger University, Saint-Louis, Sénégal\\
ba1amadou@yahoo.fr\ \ \ \ ba.amadou-diadie@ugb.edu.sn}
\date{}

\begin{document}

\maketitle

\begin{abstract}
This paper proposes a nonparametric estimation procedure for extropy and its extensions, namely the $\alpha$-Rényi and $\alpha$-Tsallis extropies, for finite discrete random variables. We establish almost sure rates of convergence and asymptotic normality for the plug-in estimators. The theoretical results are validated through a comprehensive simulation study. The findings provide a solid foundation for the use of extropy-based measures in practical applications, including forecasting, risk assessment, and decision-making under uncertainty.
\end{abstract}

\noindent \textbf{Mathematics Subject Classifications (MSC2020):} 62G05, 62G20, 62B10, 62N05, 94A17


\section{Introduction}

\subsection{Motivation}

While entropy has long been the dominant measure of uncertainty in probability and information theory, extropy offers a complementary and often more insightful perspective. Entropy emphasizes the uncertainty associated with events that occur, but it largely overlooks the role of rare or non-occurring events. Extropy, on the other hand, highlights the uncertainty of what does \emph{not} happen and is far more sensitive to low-probability outcomes. This makes it particularly valuable in fields such as risk analysis, decision theory, finance, and reliability engineering, where rare but impactful events play a crucial role.

\begin{definition}
Let $X$ be a discrete random variable defined on the probability space $(\Omega,\mathcal{A},\mathbb{P})$ with finite support $\mathcal{X} = \{c_1,c_2,\dots,c_r\}$ ($r \geq 2$) with probability mass function $p_i = \mathbb{P}(X = c_i)$ for $i \in I = \{1,\dots,r\}$.
\end{definition}

The Shannon extropy of the random variable $X$ is given by (see \cite{Lad2015})

\begin{equation}
J(\mathbf{p}) = -\sum_{i=1}^{r} (1-p_i)\log(1-p_i). \label{eq:shannon}
\end{equation}

Extropy is usually measured in bits (if $\log_2$ is used), nats (if the natural logarithm is used), or hartleys (if $\log_{10}$ is used). For ease of computation and notation, we use the natural logarithm, since logarithms of different bases are related by a constant factor.

Extropy measures the average uncertainty associated with the complement of each outcome. While entropy emphasizes diversity among likely events, extropy highlights the uncertainty embedded in the complementary probabilities.


To generalize this concept, two important one-parameter families of extropy measures have been proposed:

 \textbf{a)} The \textit{Rényi extropy} of the random variable $X$ is defined by (see \cite{Liu2023})

\begin{equation}
J_{R,\alpha}(\mathbf{p}) = \frac{1}{1-\alpha}\log\left(\sum_{i=1}^{r} (1-p_i)^\alpha\right), \quad \alpha \in (0,1)\cup(1,\infty). \label{eq:renyi}
\end{equation}

This measure is applied in tunable tail-sensitivity in information measures and decision rules with adjustable risk aversion (see \cite{Liu2023}).

\textbf{b)} The \textit{Tsallis extropy} of the random variable $X$ is defined by (see \cite{Balakrishnan2022})

\begin{equation}
J_{T,\alpha}(\mathbf{p}) = \frac{1}{1-\alpha}\left(\sum_{i=1}^{r} (1-p_i)^\alpha - 1\right), \quad \alpha \in (0,1)\cup(1,\infty). \label{eq:tsallis}
\end{equation}

This measure is used in pattern recognition, modeling non-extensive and long-tailed systems, and in robust criteria where complement probabilities matter.

As $\alpha \to 1$, both measures converge to the Shannon extropy $J(\mathbf{p})$. The parameter $\alpha$ regulates sensitivity to small or large probabilities: values $\alpha < 1$ give more weight to rare events, while $\alpha > 1$ emphasize dominant events.

From a practical perspective, these generalized extropies are powerful tools for assessing uncertainty in systems where small probabilities carry significant meaning, such as reliability analysis, risk assessment, and decision-making under uncertainty. Unlike entropy, which may underestimate uncertainty in highly concentrated distributions, extropy and its generalized forms provide a richer, complementary picture.\\

\textbf{c)}  In the study of Rényi-Tsallis extropies, a central quantity is the functional

\begin{equation}
S_\alpha(\mathbf{p}) = \sum_{i=1}^{r} (1-p_i)^\alpha, \qquad \alpha \in (0,1)\cup(1,\infty), \label{eq:S_alpha}
\end{equation}

and its empirical version

\begin{equation}
\widehat{S}_\alpha = \sum_{i=1}^{r} (1-\widehat{p}_n^i)^\alpha. \label{eq:S_alpha_emp}
\end{equation}
where $\widehat{p}_n$ denotes the empirical distribution of a multinomial sample of size $n$.
The quantity $S_{\alpha}$ appears naturally as an intermediate functional in the
definition of several generalized extropies, and its behavior governs the corresponding
statistical properties of their plug–in estimators.
The asymptotic behavior of $\widehat{S}_\alpha$  governs that of the Rényi and Tsallis extropy estimators. 

\subsection{Applications}

Extropy has emerged as a fundamental dual measure to entropy, providing a complementary perspective on uncertainty by quantifying the information contained in the unlikeliness or residual aspects of outcomes. While entropy captures the average information associated with the actualization of events, extropy focuses on the uncertainty of what does not occur, thereby offering a more balanced assessment of probabilistic systems. This duality has significant implications across disciplines.

\subsubsection{Risk Analysis and Finance}

Extropy highlights rare catastrophic outcomes often hidden in entropy-based analysis, improving robustness in portfolio management, insurance, and policy design (see \cite{Lad2015, Balakrishnan2022, Qiu2019,  Bruno2020}). Rényi/Tsallis generalizations allow tunable emphasis on the tail (when $\alpha<1$) or dominant events (when $\alpha>1)$. Recently, Rényi extropy has been applied to cryptocurrency risk analysis, demonstrating superior capability in capturing non-Gaussian dynamics compared to traditional entropy-based methods. The study by \cite{Lochaba2026} integrates Rényi extropy with machine learning models (XGBoost and k-NN) for price prediction of Bitcoin and Ethereum, showing significant improvements over conventional approaches. Additionally, \cite{Shi2025} introduced extropy-based models for portfolio optimization under uncertainty, proposing mean-variance-extropy frameworks that provide more robust investment strategies.

\subsubsection{Decision Theory}

In uncertain environments, extropy reflects the uncertainty about alternative actions not taken. Rényi and Tsallis extropy offer flexible tools for decision support under risk aversion or ambiguity preferences. Recent work by \cite{Shen2026} on partial law invariance and risk measures has explored entropic and extropic risk measures for financial decision-making under uncertainty, providing new theoretical foundations for risk-aware decision processes.

\subsubsection{Machine Learning and Artificial Intelligence}

Extropy provides valuable tools for evaluating uncertainty in classifier outputs by focusing on non-selected labels, complementing entropy-based uncertainty sampling in active learning. Recent developments include:

\begin{itemize}
    \item \cite{Jawal2022} applied residual and past versions of Tsallis and Rényi extropy to the softmax function in classification tasks, with applications to real data fitted to ARIMA models;
    \item \cite{Kharazmi2026} proposed a $\phi$-extropy divergence measure for binary classification via optimal thresholding. The method was compared with ten standard machine learning algorithms (logistic regression, SVM, random forest, k-NN, Naive Bayes, LDA, QDA, AdaBoost, gradient boosting, and decision trees) on heart disease detection data, with evaluation metrics including precision, recall, and F1-score;
    \item \cite{Kumar2025} developed an extropy rate-based method for feature selection, demonstrating competitive performance against established methods such as entropy rate and other state-of-the-art variable selection techniques;
    \item \cite{Mahesh2026} released the \texttt{RenyiExtropy} package for R, providing computational tools for entropy and extropy measures (Shannon, Rényi, Tsallis) for discrete probability distributions, with applications in information theory, statistics, and machine learning.
\end{itemize}

\subsubsection{Reliability and Survival Analysis}

Cumulative residual extropy and its dynamic extensions provide refined tools for modeling system lifetimes and medical survival data. \cite{DiCrescenzo2019} studied cumulative entropies and extropies for lifetime distributions, while \cite{Asadi2018} introduced dynamic cumulative residual extropy with applications in reliability engineering. These measures are particularly useful for assessing the remaining uncertainty in aging systems and medical prognosis.

\subsubsection{Pattern Recognition}

\cite{Balakrishnan2022} demonstrated the application of Tsallis extropy as a measure of discrimination in pattern recognition tasks, establishing its properties, bounds, and comparative advantages over traditional methods.

\subsubsection{Health Data Analysis}

\cite{Aldallal2025} explored weighted Tsallis extropy applied to human health data, allowing emphasis on rare but severe health events. The weighting mechanism enables practitioners to focus on critical health outcomes, making it particularly valuable in epidemiological studies and clinical decision-making.

\subsubsection{Information Theory and Statistics}

Extropy complements entropy-based measures in coding, inference, and divergence analysis, often yielding advantages in small-sample or imbalanced data scenarios (see \cite{Hood2015, Qiu2018}). \cite{Qiu2018} introduced nonparametric estimators for extropy with applications to testing uniformity, while \cite{Jahanshahi2020} proposed kernel-based density estimators for extropy, investigating their asymptotic properties including consistency.

\subsection{Previous Work on Estimation}

Various estimation techniques have been developed to quantify extropy in different contexts:

\begin{itemize}
    \item \cite{Qiu2018} introduced nonparametric estimators for extropy with applications in testing uniformity, establishing a foundation for frequency-based estimation;
    \item \cite{Jahanshahi2020} and \cite{Alizadeh2020} proposed kernel-based density estimators for extropy. The former investigated the asymptotic properties (consistency) of the estimator, while the latter compared the performance of kernel-based estimators with other nonparametric methods;
    \item \cite{AlLabadi2020} proposed a Bayesian approach based on the Dirichlet process for the estimation of extropy, providing a fully probabilistic framework for uncertainty quantification;
    \item \cite{Tahmasebi2020} introduced negative cumulative extropy with applications in various statistical problems, including compressive sensing \cite{Kazemi2021};
    \item Recent work by \cite{Lochaba2026} and \cite{Kharazmi2026} has integrated extropy measures with machine learning frameworks, opening new avenues for data-driven uncertainty quantification.
\end{itemize}

Despite this growing body of work, to our knowledge, no results regarding the almost sure consistency and the asymptotic normality of Shannon, Rényi, and Tsallis extropy estimators were known prior to this work. Our contribution fills this gap by providing a unified asymptotic theory for these important measures.

\subsection{Main Contribution}

Our main contribution may be summarized as follows. Given an i.i.d. sample $X_1,\dots,X_n$ from $X$ with distribution $\mathbf{p}$, we derive almost sure convergence and central limit theorems for Shannon, $\alpha$-Rényi, and $\alpha$-Tsallis extropies.

Our method consists of first obtaining general laws for an arbitrary functional of the form

\begin{equation}
F(\mathbf{p}) = \sum_{i\in I} \phi(p_i), \label{eq:phi_func}
\end{equation}

where $\phi: (0,1) \to \mathbb{R}$ is a twice continuously differentiable function. \\
The results on the functional $F(\mathbf{p})$, which is also known as the functional $\phi$-extropy, will lead to those of the extropies mentioned above.

\subsection{Overview of the Paper}

The remainder of the paper is organized as follows. In Section \ref{section2}, we present the distribution limit for the empirical functional $\phi$-extropy. Section \ref{section3} derives the asymptotic limit laws for Shannon, Rényi, and Tsallis extropies. Section \ref{section4} contains the proofs of the main results. Section \ref{section5} presents a comprehensive simulation study, including convergence properties and an application to forecasting. Finally, Section \ref{section6} concludes the paper and discusses future research directions.

\section{Distribution Limit for Empirical Functional $\phi$-Extropy}
\label{section2}
\subsection{Notation and Main Results}

Let $X$ be a random variable defined on the probability space $(\Omega,\mathcal{A},\mathbb{P})$ taking values in $\mathcal{X} = \{c_1,c_2,\dots,c_r\}$, with probability distribution $\mathbf{p} = (p_i)_{1\leq i\leq r}$ such that for any $i \in I = \{1,2,\dots,r\}$,

\begin{equation}
p_i = \mathbb{P}(X = c_i). \label{eq:prob}
\end{equation}

In general, the full probability distribution $\mathbf{p} = (p_i)_{1\leq i\leq r}$ is unknown. Let $X_1,\dots,X_n$ be $n$ i.i.d. copies of $X$ according to $\mathbf{p}$. For a given $i \in I$, define the easiest and most objective estimator of $p_i$ and based on the  i.i.d. sample $X_1,\dots,X_n$ be $n$ :

\begin{equation}
\widehat{p}_n^i = \frac{1}{n}\sum_{j=1}^{n} \mathbf{1}_{c_i}(X_j), \label{eq:empirical}
\end{equation}

where
\[
\mathbf{1}_{c_i}(X_j) =
\begin{cases}
1 & \text{if } X_j = c_i,\\
0 & \text{otherwise}.
\end{cases}
\]

For a fixed $i\in I$, this empirical estimator $\widehat{p}_n^i$ of $p_i$ is strongly consistent and asymptotically normal. Precisely, when $n$ tends to infinity,

\begin{equation}
\widehat{p}_n^i - p_i\stackrel{a.s.}{ \rightarrow} 0, \label{eq:strong_consistency}
\end{equation}

\begin{equation}
\sqrt{n}(\widehat{p}_n^i - p_i) \stackrel{\mathcal{D}}{ \rightsquigarrow} \mathcal{N}(0, p_i(1-p_i)), \label{eq:asymptotic_normality}
\end{equation}

where symbols $\stackrel{a.s.}{ \rightarrow}$ and  $\stackrel{\mathcal{D}}{ \rightsquigarrow}
$ denote respectively \textit{almost sure convergence} and \textit{convergence in distribution}.

 \noindent These asymptotic properties derive from the law of large
numbers and central limit theorem.
\\ The corresponding plug-in estimators of the \textit{r.v} $X$ are obtained by replacing $\mathbf{p}$ by $\widehat{\mathbf{p}}_n$: 
\begin{eqnarray}
 \label{extroplug} J(\widehat{p}_{n})&=&-\sum_{i=1}^r(1-\widehat{p}_n^i)\log(1-\widehat{p}_n^i)\\
 \label{reyniplug} J_{R,\alpha}(\widehat{p}_{n})&=&\frac{1}{1-\alpha}\log \left( \sum_{i=1}^r(1-\widehat{p}_n^i)^\alpha\right)\\ \text{and}\ \
 \label{tsallisplug} J_{T,\alpha}(\widehat{p}_{n})&=&\frac{1}{1-\alpha}\left(\sum_{i=1}^r (1-\widehat{p}_n^i)^\alpha-1\right)
\end{eqnarray}

\begin{remark}
Theses plug-in estimators are biased in finite samples. However, when $n$ is large, 
 the bias is of order $O(n^{-1})$, and the bias-corrected versions differ from the uncorrected ones only by terms of order $O(n^{-1})$. Consequently, the uncorrected and corrected estimators are asymptotically equivalent. Therefore, all asymptotic results presented in this paper (consistency, asymptotic normality, and confidence intervals) apply equally to both versions. For a detailled study of 
 finite-sample bias correction, we refer the reader to the companion paper \cite{Ba2026b}.
\end{remark}

\subsection{Functional $\phi$-Extropy}

\begin{definition}
Let $\phi: (0,1) \to \mathbb{R}$ be a twice continuously differentiable function. The functional $\phi$-extropy of the probability distribution $\mathbf{p} = (p_i)_{i\in I}$ is given by

\begin{equation}
F(\mathbf{p}) = \sum_{i\in I} \phi(p_i). \label{eq:F_def}
\end{equation}
\end{definition}

This functional encompasses the extropy measures of interest as special cases:
\begin{itemize}
\item For $\phi(s) = -(1-s)\log(1-s)$, we obtain the Shannon extropy $J(\mathbf{p})$ (see \eqref{eq:shannon});
\item For $\phi(s) = (1-s)^\alpha$, we obtain the functional $S_\alpha(\mathbf{p})$ from which Rényi and Tsallis extropies are derived (see \eqref{eq:renyi} and \eqref{eq:tsallis}).
\end{itemize}

Thus, the results obtained for the functional $F(\mathbf{p})$ will directly apply to the particular cases of Shannon, Rényi, and Tsallis extropies.

Based on \eqref{eq:empirical}, we define the empirical functional $\phi$-extropy:

\begin{equation}
F(\widehat{\mathbf{p}}_n) = \sum_{i\in I} \phi(\widehat{p}_n^i). \label{eq:F_emp}
\end{equation}

\subsection{Main Theorem}

We now state the main result concerning the almost sure convergence and asymptotic normality of the empirical functional $\phi$-extropy $F(\widehat{\mathbf{p}}_n)$ (see \eqref{eq:F_emp}).

Define

\begin{eqnarray}
A_F(\mathbf{p}) &= &\sum_{i\in I} |\phi'(p_i)|, \label{eq:A_F}\\
\nonumber \text{and}\ \ \ \ \ \ \ \ \ \ &&\\
\sigma_F^2(\mathbf{p}) &=& \sum_{i\in I} p_i(1-p_i)(\phi'(p_i))^2 - 2\sum_{(i,j)\in I^2, i\neq j} (p_i p_j)^{3/2} \phi'(p_i)\phi'(p_j). \label{eq:sigma_F}
\end{eqnarray}

\begin{theorem}
\label{thm:main}
Let $\mathbf{p} = (p_i)_{i\in I}$ be a probability distribution and $\widehat{\mathbf{p}}_n = (\widehat{p}_n^i)_{i\in I}$ be generated by an i.i.d. sample $X_1,\dots,X_n$ from $X$ according to $\mathbf{p}$. Then the following asymptotic results hold:

\begin{eqnarray}
&&\limsup_{n\to\infty} \frac{|F(\widehat{\mathbf{p}}_n) - F(\mathbf{p})|}{a_n} \leq A_F(\mathbf{p}) \quad \text{a.s.}, \label{eq:almost_sure_F}\\
&&\sqrt{n}(F(\widehat{\mathbf{p}}_n) - F(\mathbf{p})) \stackrel{\mathcal{D}}{ \rightsquigarrow} \mathcal{N}(0, \sigma_F^2(\mathbf{p})) \quad \text{as } n \to +\infty, \label{eq:normal_F}
\end{eqnarray}

where $a_n = \sup_{i\in I} |\widehat{p}_n^i - p_i|$.
\end{theorem}


\section{Extropies Asymptotic Limit Laws}
\label{section3}
\subsection{Asymptotic Behavior of $S_\alpha(\widehat{\mathbf{p}}_n)$} 
In the following corollary, we establish the almost sure convergence and the asymptotic
normality of the estimator $\widehat{S}_\alpha(\widehat{\mathbf{p}}_n)$. \\
Let $\widehat{S}_\alpha(\widehat{\mathbf{p}}_n)$ defined by \eqref{eq:S_alpha_emp}  and for $\alpha \in (0,1) \cup (1,\infty)$, set 

\begin{eqnarray}
A_{S_\alpha}(\mathbf{p}) & = &\alpha \sum_{i\in I} (1-p_i)^{\alpha-1} \label{eq:A_Salpha}\\
\nonumber \text{and}\ \ \ \ \ \ \ \ \ &&\\
\sigma_{S_\alpha}^2(\mathbf{p}) &=& \alpha^2 \Biggr( \sum_{i\in I} p_i(1-p_i)^{2\alpha-1} - 2\sum_{(i,j)\in I^2, i\neq j} (p_i p_j)^{3/2} (1-p_i)^{\alpha-1}(1-p_j)^{\alpha-1} \Biggr). \label{eq:sigma_Salpha}
\end{eqnarray}

\begin{corollary}
\label{cor:S_alpha}
Under the same assumptions as in Theorem \ref{thm:main} the following hold:

\begin{eqnarray}
&& \limsup_{n\to\infty} \frac{|S_\alpha(\widehat{\mathbf{p}}_n) - S_\alpha(\mathbf{p})|}{a_n} \leq A_{S_\alpha}(\mathbf{p}) \quad \text{a.s.}, \label{eq:almost_sure_Salpha}\\
&& \sqrt{n}(S_\alpha(\widehat{\mathbf{p}}_n) - S_\alpha(\mathbf{p})) \stackrel{\mathcal{D}}{ \rightsquigarrow} \mathcal{N}(0, \sigma_{S_\alpha}^2(\mathbf{p})) \quad \text{as } n \to +\infty. \label{eq:normal_Salpha}
\end{eqnarray}
\end{corollary}

\subsection{Asymptotic Behavior of the Shannon Extropy Estimator.}

The Corollary \ref{cor:J}  below establishes the asymptotic behavior of the Shannon extropy estimator. 
\noindent Let $ J(\widehat{p}_{n})$ define by \eqref{extroplug} and set
\begin{eqnarray}
A_J(\mathbf{p}) &=& \sum_{i\in I} |1 + \log(1-p_i)|, \label{eq:A_J}\\
\notag \text{and} \ \ \ \qquad && \\
\notag  \sigma_J^2(\mathbf{p}) &=& \sum_{i\in I} p_i(1-p_i)(1 + \log(1-p_i))^2 \\
&&\ \ \qquad \ \ - 2\sum_{(i,j)\in I^2, i\neq j} (p_i p_j)^{3/2} (1 + \log(1-p_i))(1 + \log(1-p_j)). \label{eq:sigma_J}
\end{eqnarray}

\begin{corollary}
\label{cor:J}
Under the same assumptions as in Theorem \ref{thm:main},  the following
asymptotic results hold :
\begin{eqnarray}
&& \limsup_{n\to\infty} \frac{|J(\widehat{\mathbf{p}}_n) - J(\mathbf{p})|}{a_n} \leq A_J(\mathbf{p}) \quad \text{a.s.}, \label{eq:almost_sure_J}
\\
&&
\sqrt{n}(J(\widehat{\mathbf{p}}_n) - J(\mathbf{p})) \stackrel{\mathcal{D}}{ \rightsquigarrow} \mathcal{N}(0, \sigma_J^2(\mathbf{p})) \quad \text{as } n \to +\infty. \label{eq:normal_J}
\end{eqnarray}
\end{corollary}

\subsection{Asymptotic Behavior of the Rényi Extropy Estimator}
The Corollary \ref{cor:R}  below establishes the asymptotic behavior of the Rényi extropy estimator. 
Let $ J_{R,\alpha}(\widehat{p}_{n})$ defined by \eqref{reyniplug} and for $\alpha \in (0,1) \cup (1,\infty)$ set
\begin{eqnarray}
A_{R,\alpha}(\mathbf{p}) & = &\frac{\alpha}{|1-\alpha| S_\alpha(\mathbf{p})} \sum_{i\in I} (1-p_i)^{\alpha-1} \label{eq:A_R}\\
\notag \text{and}&& \ \ \ \ \ \ \\
\notag \sigma_{R,\alpha}^2(\mathbf{p}) &=& \left(\frac{\alpha}{(1-\alpha)S_\alpha(\mathbf{p})}\right)^2 \Biggr( \sum_{i\in I} p_i(1-p_i)^{2\alpha-1} \\
&&\ \ \qquad \ \  \qquad - 2\sum_{(i,j)\in I^2, i\neq j} (p_i p_j)^{3/2} (1-p_i)^{\alpha-1}(1-p_j)^{\alpha-1} \Biggr). \label{eq:sigma_R}
\end{eqnarray}

\begin{corollary}
\label{cor:R}
Under the same assumptions as in Theorem \ref{thm:main}, the following
asymptotic results hold :
\begin{eqnarray}
&&\limsup_{n\to\infty} \frac{|J_{R,\alpha}(\widehat{\mathbf{p}}_n) - J_{R,\alpha}(\mathbf{p})|}{a_n} \leq A_{R,\alpha}(\mathbf{p}) \quad \text{a.s.}, \label{eq:almost_sure_R}\\
&&\sqrt{n}(J_{R,\alpha}(\widehat{\mathbf{p}}_n) - J_{R,\alpha}(\mathbf{p})) \stackrel{\mathcal{D}}{ \rightsquigarrow} \mathcal{N}(0, \sigma_{R,\alpha}^2(\mathbf{p})) \quad \text{as } n \to +\infty. \label{eq:normal_R}
\end{eqnarray}
\end{corollary}

\subsection{Asymptotic Behavior of the Tsallis Extropy Estimator}
The Corollary \ref{cor:T}  below establishes the asymptotic behavior of the Tsallis extropy estimator. Let $J_{T,\alpha}(\widehat{p}_{n})$ be defined by \eqref{tsallisplug} and for $\alpha \in (0,1) \cup (1,\infty)$ set  
\begin{eqnarray}
A_{T,\alpha}(\mathbf{p}) &=& \frac{\alpha}{|\alpha-1|} \sum_{i\in I} (1-p_i)^{\alpha-1} \label{eq:A_T}\\
\text{and} \notag \ \ \ \ \ \ \  &&\\
\notag \sigma_{T,\alpha}^2(\mathbf{p}) &=& \left(\frac{\alpha}{\alpha-1}\right)^2 \Biggr( \sum_{i\in I} p_i(1-p_i)^{2\alpha-1}\\
  &&  \ \ \  \ \ \ \ \ \ \ \ \ \ \ \ \ \ \ \ \ \  - 2\sum_{(i,j)\in I^2, i\neq j} (p_i p_j)^{3/2} (1-p_i)^{\alpha-1}(1-p_j)^{\alpha-1} \Biggr). \label{eq:sigma_T}
\end{eqnarray}

\begin{corollary}
\label{cor:T}
Under the same assumptions as in Theorem \ref{thm:main}, the following
asymptotic results hold :

\begin{eqnarray}
&& \limsup_{n\to\infty} \frac{|J_{T,\alpha}(\widehat{\mathbf{p}}_n) - J_{T,\alpha}(\mathbf{p})|}{a_n} \leq A_{T,\alpha}(\mathbf{p}) \quad \text{a.s.}, \label{eq:almost_sure_T}\\
&& 
\sqrt{n}(J_{T,\alpha}(\widehat{\mathbf{p}}_n) - J_{T,\alpha}(\mathbf{p})) \stackrel{\mathcal{D}}{ \rightsquigarrow} \mathcal{N}(0, \sigma_{T,\alpha}^2(\mathbf{p})) \quad \text{as } n \to +\infty. \label{eq:normal_T}
\end{eqnarray}
\end{corollary}

\section{Proofs}
\label{section4}
Before presenting the proofs, we introduce the following notations. For a fixed $i \in I$, denote

\[
\Delta_n^i = \widehat{p}_n^i - p_i, \qquad \delta_n(p_i) = \sqrt{\frac{n}{p_i}} \Delta_n^i,
\]

and

\[
a_n = \sup_{i\in I} |\Delta_n^i|.
\]

We recall that, since for a fixed $i \in I$, $n\widehat{p}_n^i$ has a binomial distribution with parameters $n$ and success probability $p_i$, we have

\[
\mathbb{E}[\widehat{p}_n^i] = p_i \quad \text{and} \quad \operatorname{Var}(\widehat{p}_n^i) = \frac{p_i(1-p_i)}{n}.
\]

By the asymptotic Gaussian limit of the multinomial law (see, for example, \cite{Lo2016}), we have

\begin{equation}
(\delta_n(p_i), i \in I) \stackrel{\mathcal{D}}{ \rightsquigarrow} Z(\mathbf{p}) \sim \mathcal{N}(0, \Sigma_{\mathbf{p}}) \quad \text{as } n \to +\infty, \label{eq:multinomial_limit}
\end{equation}

\noindent where $Z(\mathbf{p}) = (Z_{p_i}, i \in I)^T$ is a centered Gaussian random vector of dimension $\#(I)=r$ with covariance elements
\begin{equation}
(\Sigma_{\mathbf{p}})_{(i,j)} = (1-p_i)\delta_{i,j} - \sqrt{p_i p_j} (1-\delta_{i,j}), \label{eq:covariance}
\end{equation}
where $\delta_{i,j} = \begin{cases} 1 \ \ \ \ \ \text{   if   }\ \ i = j\\
0\ \  \text{  otherwise }\end{cases}$ where $(i,j)\in I^2$

\subsection{Proof of Theorem \ref{thm:main}}
Let $\phi: (0,1) \to \mathbb{R}$ be a twice continuously differentiable function. Then for fixed $i \in I$, we have

\begin{equation}
\phi(\widehat{p}_n^i) = \phi(p_i + \Delta_n^i) = \phi(p_i) + \Delta_n^i \phi'(p_i + \theta_1(i)\Delta_n^i), \label{eq:phi_expansion1}
\end{equation}

by the mean value theorem applied to $\phi$, where $\theta_1(i) $  is some number lying in $(0,1)$.

Applying the mean value theorem again to $\phi'$:

\begin{equation}
\phi'(p_i + \theta_1(i)\Delta_n^i) = \phi'(p_i) + \theta_1(i)\Delta_n^i \phi''(p_i + \theta_2(i)\Delta_n^i), \label{eq:phi_expansion2}
\end{equation}
where $\theta_2(i)$  is some number lying in $(0,1)$.

We can write \eqref{eq:phi_expansion1} as

\begin{equation}
\phi(\widehat{p}_n^i) = \phi(p_i) + \Delta_n^i \phi'(p_i) + \theta_1(i)(\Delta_n^i)^2 \phi''(p_i + \theta_2(i)\Delta_n^i). \label{eq:phi_expansion3}
\end{equation}
Now we have by summation over $i \in I$:

\begin{equation}
F(\widehat{\mathbf{p}}_n) - F(\mathbf{p}) = \sum_{i\in I} \Delta_n^i \phi'(p_i) + \sum_{i\in I} \theta_1(i)(\Delta_n^i)^2 \phi''(p_i + \theta_2(i)\Delta_n^i). \label{eq:F_expansion}
\end{equation}

Hence

\[
|F(\widehat{\mathbf{p}}_n) - F(\mathbf{p})| \leq a_n \sum_{i\in I} |\phi'(p_i)| + a_n^2 \sum_{i\in I} |\phi''(p_i + \theta_2(i)\Delta_n^i)|.
\]

Therefore,

\begin{equation}
\limsup_{n\to\infty} \frac{|F(\widehat{\mathbf{p}}_n) - F(\mathbf{p})|}{a_n} \leq A_F(\mathbf{p}) \quad \text{a.s.}, \label{eq:F_almost_sure_proof}
\end{equation}

since $a_n \xrightarrow{\text{a.s.}} 0$ as $n \to +\infty$ and

\[
\sum_{i\in I} |\phi''(p_i + \theta_2(i)\Delta_n^i)| \to \sum_{i\in I} |\phi''(p_i)| < \infty.
\]

This proves \eqref{eq:almost_sure_F}.

Now we prove \eqref{eq:normal_F}. From \eqref{eq:F_expansion}, we obtain 
\begin{equation}
\sqrt{n}(F(\widehat{\mathbf{p}}_n) - F(\mathbf{p})) = \sum_{i\in I} \sqrt{p_i} \delta_n(p_i) \phi'(p_i) + \sqrt{n} R_n, \label{eq:F_normal_expansion}
\end{equation}
where
\[
R_n = \sum_{i\in I} \theta_1(i)(\Delta_n^i)^2 \phi''(p_i + \theta_2(i)\Delta_n^i).
\]
Using \eqref{eq:multinomial_limit}, we obtain
\begin{equation}
\sum_{i\in I} \sqrt{p_i} \delta_n(p_i) \phi'(p_i) \stackrel{\mathcal{D}}{ \rightsquigarrow} \sum_{i\in I} \phi'(p_i) \sqrt{p_i} Z_{p_i} \quad \text{as } n \to +\infty, \label{eq:normal_limit}
\end{equation}

which follows a centered normal law with variance $\sigma_F^2(\mathbf{p})$ where 
$$\sigma_F^2(\mathbf{p}) = \sum_{i\in I} p_i(1-p_i)(\phi'(p_i))^2 - 2\sum_{(i,j)\in I^2, i\neq j} (p_i p_j)^{3/2} \phi'(p_i)\phi'(p_j)$$
 since

\begin{eqnarray}
\notag \operatorname{Var}\left(\sum_{i\in I} \sqrt{p_i} \phi'(p_i) Z_{p_i}\right)
&=&\sum_{i\in I}\operatorname{Var}\left(  \phi'(p_i) \sqrt{p_i}Z_{p_i}\right)+ 2\sum_{(i,j)\in I^2, i\neq j} Cov\left(  \phi'(p_i) \sqrt{p_i}Z_{p_i},\phi'(p_j) \sqrt{p_j}Z_{p_j}\right) \\
&=& \sum_{i\in I} p_i (1-p_i)(\phi'(p_i))^2 - 2\sum_{(i,j)\in I^2, i\neq j} p_i p_j \sqrt{p_ip_j} \phi'(p_i)\phi'(p_j).
\end{eqnarray}

The proof will be complete if we show that $\sqrt{n} R_n$ converges to zero in probability.

We have

\[
|\sqrt{n} R_n| \leq \sqrt{n} a_n^2 \sum_{i\in I} |\phi''(p_i + \theta_2(i)\Delta_n^i)|.
\]

By the Bienaymé-Chebyshev inequality, for any $\epsilon > 0$ and for fixed $i \in I$,

\[
\mathbb{P}(\sqrt{n}(\widehat{p}_n^i - p_i)^2 \geq \epsilon) = \mathbb{P}\left(|\widehat{p}_n^i - p_i| \geq \frac{\sqrt{\epsilon}}{n^{1/4}}\right) \leq \frac{p_i(1-p_i)}{\epsilon n^{1/2}}.
\]

Hence $\sqrt{n} a_n^2 = o_p(1)$ since $ \sum_{i\in I} |\phi''(p_i + \theta_2(i)\Delta_n^i)| <\infty $.\\
 This proves \eqref{eq:normal_F} and ends the proof of Theorem \ref{thm:main}.

\subsection{Proofs of Corollaries}
\textbf{A-} The proofs of Corollaries \ref{cor:S_alpha} and \ref{cor:J} are direct adaptation of Theorem \ref{thm:main} with respectively $\phi(s) = (1-s)^\alpha$ and  $\phi(s) = -(1-s)\log(1-s)$. \\

%

\textbf{B-}
Proof of Corollary \ref{cor:R}. For $\alpha \in (0,1) \cup (1,\infty)$, the $\alpha$-Rényi extropy is expressed through $S_\alpha(\mathbf{p}) = \sum_{i\in I} \phi(p_i)$ with $\phi(s) = (1-s)^\alpha$.

We have
\begin{eqnarray}
\notag J_{R,\alpha}(\widehat{\mathbf{p}}_n) - J_{R,\alpha}(\mathbf{p}) &=& \frac{1}{1-\alpha} \left( \log S_\alpha(\widehat{\mathbf{p}}_n) - \log S_\alpha(\mathbf{p}) \right)\\
\notag &=& \frac{1}{1-\alpha}\log \left(1+ \frac{ S_\alpha(\widehat{\mathbf{p}}_n)-S_\alpha(\mathbf{p}) }{S_\alpha(\mathbf{p}) }\right).
\end{eqnarray}
Using a Taylor expansion of $\log(1+y)$, it follows that, almost surely,
\begin{equation}
J_{R,\alpha}(\widehat{\mathbf{p}}_n) - J_{R,\alpha}(\mathbf{p}) =\frac{1}{(1-\alpha)S_\alpha(\mathbf{p})}\left( S_\alpha(\widehat{\mathbf{p}}_n)-S_\alpha(\mathbf{p}) \right).\label{jra}
\end{equation} 
This, combined with \eqref{eq:almost_sure_Salpha} of Corollary \ref{cor:S_alpha}, proves \eqref{eq:almost_sure_R}. \\

\noindent Now from \eqref{eq:F_normal_expansion}, we obtain
\[
\sqrt{n}(S_\alpha(\widehat{\mathbf{p}}_n) - S_\alpha(\mathbf{p})) = \sqrt{n} \sum_{i\in I} \Delta_n^i \phi'(p_i) + o_p(1),
\]

where $\phi'(p_i) = -\alpha(1-p_i)^{\alpha-1}$.

Hence, dividing each member by $\sqrt{n}S_\alpha(\textbf{p})$ gives

\[
\frac{S_\alpha(\widehat{\mathbf{p}}_n)}{S_\alpha(\mathbf{p})} = 1 + \frac{\sum_{i\in I} \Delta_n^i \phi'(p_i)}{S_\alpha(\mathbf{p})} + o_p(1).
\]

By Taylor expansion of $\log(1+y)$, it follows that, almost surely 

\[
\log S_\alpha(\widehat{\mathbf{p}}_n) - \log S_\alpha(\mathbf{p}) = \frac{\sum_{i\in I} \Delta_n^i \phi'(p_i)}{S_\alpha(\mathbf{p})} + O_p\left(\frac{1}{n}\right).
\]
but 
 \begin{eqnarray*}
 \sum_{i\in I}  \sqrt{p_i}\delta_n(p_i) \phi'( p_{i})&\stackrel{\mathcal{D}}{ \rightsquigarrow}& \sum_{i\in I}\  \phi'( p_{i})\sqrt{p_i}Z_{p_i}
,\ \ \text{as}\ \ n\rightarrow+\infty,\\
&\stackrel{\mathcal{D}}{ \rightsquigarrow} & 
\mathcal{N}\left( 0,\sigma _{R}^{2}(\textbf{p})\right)\text{ as } n\rightarrow +\infty,
 \end{eqnarray*}
where \begin{eqnarray*}
\sigma _{R}^{2}(\textbf{p})&=&\sum_{i\in I} p_i(1-p_i)( \phi'( p_{i}))^2-2\sum_{(i,j)\in I^2,i\neq j}(p_ip_j)^{3/2}\phi'( p_{i} )\phi'( p_{j}),
\end{eqnarray*}using \eqref{eq:normal_limit}. Finally
\begin{eqnarray*}
\sqrt{n}\left( J_{R,\alpha}( \widehat{\textbf{p}}_n )-J_{R,\alpha }(\textbf{p})\right) &\stackrel{\mathcal{D}}{ \rightsquigarrow} & 
\mathcal{N}\left( 0,\sigma _{R,\alpha}^{2}(\textbf{p})\right)\text{ as } n\rightarrow +\infty,
\end{eqnarray*}
where 
 $$ \sigma _{R,\alpha}^{2}(\textbf{p})=\left(\frac{\alpha}{(1-\alpha)\mathcal{S}_\alpha(\textbf{p})}\right)^2\left(\sum_{i\in I} p_i(1-p_i)^{2\alpha-1}-2\sum_{(i,j)\in I^2,i\neq j}(p_i p_j)^{3/2}(1-p_i)^{\alpha-1}(1- p_{j})^{\alpha-1}\right)$$

\noindent This proves \eqref{eq:normal_R} and ends the proof of the Corollary \ref{cor:R}.\\

\bigskip \textbf{C-} Proof of Corollary \ref{cor:T}. Since the Tsallis extropy is related to $S_\alpha(\mathbf{p})$ through a linear transformation, the proof follows directly from  \eqref{eq:almost_sure_Salpha} and \eqref{eq:normal_Salpha} 
.

\section{Simulation Study}
\label{section5}
\subsection{Motivation}

In this section, we conduct a comprehensive simulation study to:
\begin{enumerate}
    \item validate the asymptotic theory established in Section \ref{section3}, namely the almost sure convergence and asymptotic normality of the plug-in estimators;
    \item illustrate the practical behavior of the estimators for finite sample sizes;
    \item demonstrate the relevance of extropy-based measures in a forecasting context.
\end{enumerate}

\subsection{Experimental Setup}

\noindent We generate $M = 1000$ independent samples of sizes 
$n = 10, 20, 50, 100, 200, 500, 1000, 2000$ from a multinomial distribution with $r = 5$ categories and probability vector
\[
\mathbf{p} = (0.40, 0.30, 0.15, 0.10, 0.05).
\]
This distribution represents a realistic scenario with one dominant category (40\%) and a tail of decreasing probabilities, mimicking many real-world forecasting situations.

\noindent For each sample, we compute the following plug-in estimators:
\begin{itemize}
    \item The Shannon extropy: $\displaystyle \widehat{J} = -\sum_{i=1}^{r} (1-\widehat{p}_i)\log(1-\widehat{p}_i)$ (see \eqref{eq:shannon});
    \item The Rényi extropy for $\alpha = 0.5$ and $\alpha = 3$:
    \[
    \widehat{J}_{R,\alpha} = \frac{1}{1-\alpha}\log\left(\sum_{i=1}^{r} (1-\widehat{p}_i)^\alpha\right)
    \]
    (see \eqref{eq:renyi});
    \item The Tsallis extropy for $\alpha = 0.5$ and $\alpha = 3$:
    \[
    \widehat{J}_{T,\alpha} = \frac{1}{1-\alpha}\left(\sum_{i=1}^{r} (1-\widehat{p}_i)^\alpha - 1\right)
    \]
    (see \eqref{eq:tsallis}).
\end{itemize}

\noindent The true values $J$, $J_{R,\alpha}$, and $J_{T,\alpha}$ are computed using the true probability vector $\mathbf{p}$.

\subsection{Shannon Extropy: Results}
\begin{figure}[h!]
\centering
\includegraphics[scale=0.36]{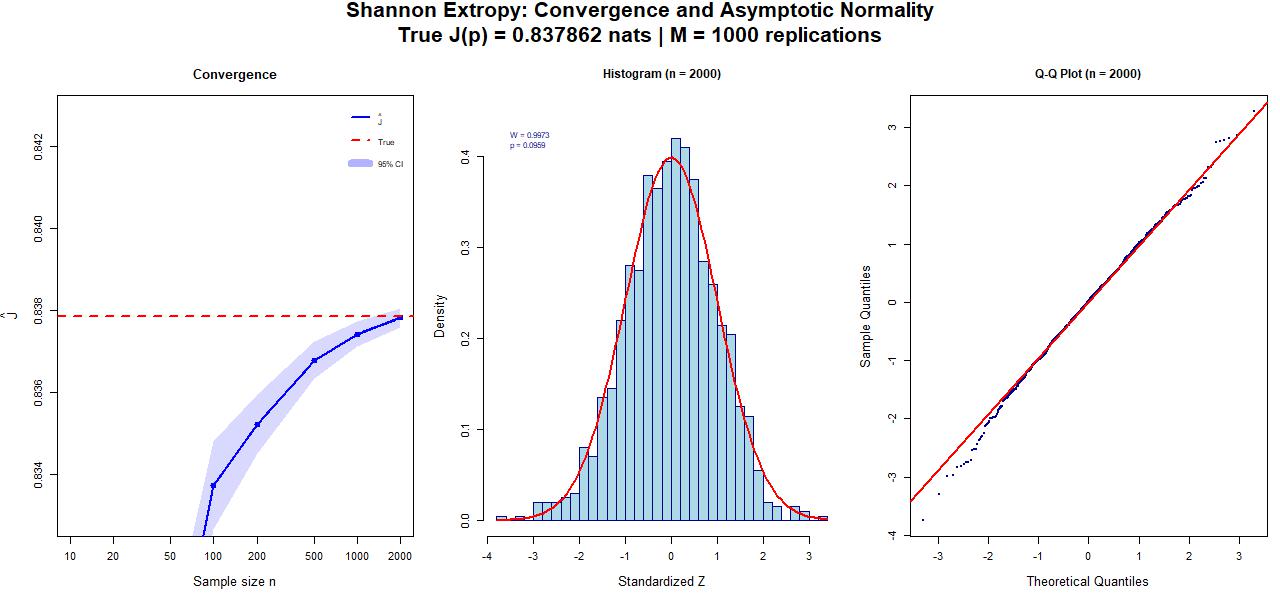} 
\caption{Shannon extropy estimator. \textbf{(a)} Convergence of $\widehat{J}$ to $J(\mathbf{p})$ (red dashed line) with 95\% confidence band. \textbf{(b)} Histogram of standardized estimates for $n = 2000$ (overlaid with standard normal density). \textbf{(c)} Q-Q plot for $n = 2000$. 
The simulation results provide empirical support for the theoretical convergence and asymptotic normality results.}
\label{fig:shannon}
\end{figure}

\noindent \textsc{Figure} \ref{fig:shannon} presents the convergence and asymptotic normality of the Shannon extropy estimator. Three panels are displayed in a single row:

\begin{itemize}
    \item \textbf{Panel (a):} Convergence of $\widehat{J}$ to the true value $J(\mathbf{p})$ with 95\% confidence bands. As $n$ increases, the estimate stabilizes around the true value and the confidence bands shrink, confirming the almost sure convergence established in Corollary \ref{cor:J}.
    \item \textbf{Panel (b):} Histogram of the standardized estimates for $n = 2000$, overlaid with the standard normal density. The excellent agreement confirms the asymptotic normality result.
    \item \textbf{Panel (c):} Q-Q plot for $n = 2000$. The alignment of points with the diagonal line provides further visual confirmation of normality.
\end{itemize}

\subsection{Rényi Extropy: Results}

\begin{figure}[h!]
\centering
\includegraphics[scale=0.36]{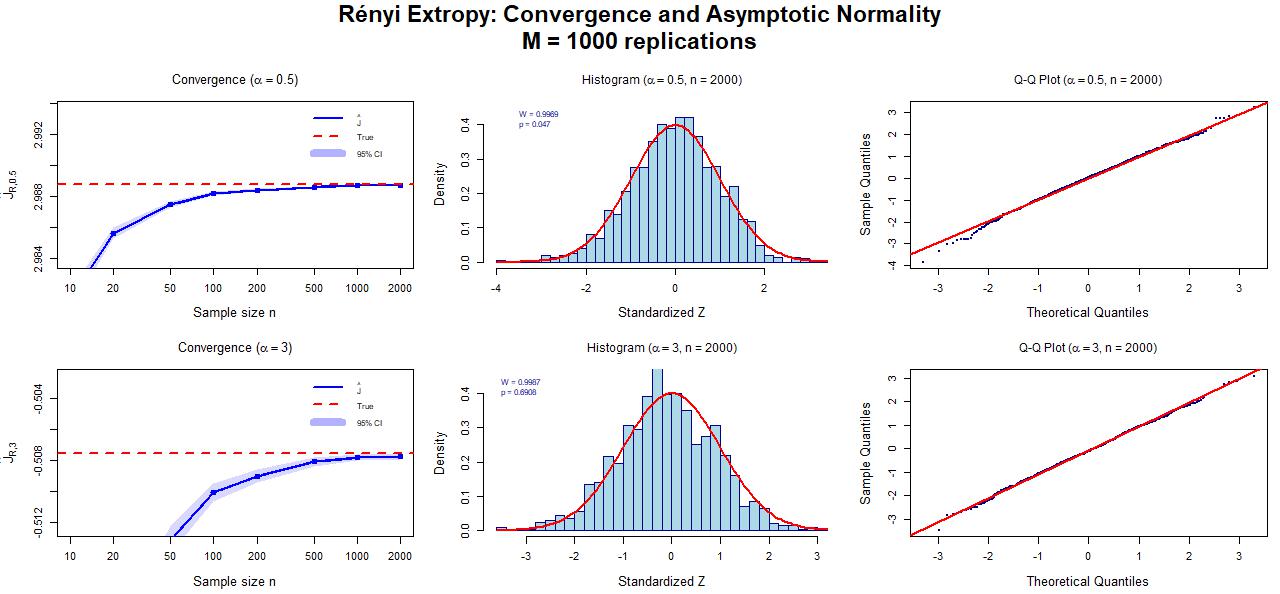} 
\caption{Rényi extropy estimator. Top row ($\alpha = 0.5$): \textbf{(a)} Convergence of $\widehat{J}_{R,0.5}$ to $J_{R,0.5}$ with 95\% confidence band. \textbf{(b)} Histogram of standardized estimates for $n = 2000$. \textbf{(c)} Q-Q plot for $n = 2000$. Bottom row ($\alpha = 3$): \textbf{(d)} Convergence of $\widehat{J}_{R,3}$ to $J_{R,3}$ with 95\% confidence band. \textbf{(e)} Histogram of standardized estimates for $n = 2000$. \textbf{(f)} Q-Q plot for $n = 2000$. The simulation results provide empirical support for the theoretical convergence and asymptotic normality results for both values of $\alpha$.}
\label{fig:renyi}
\end{figure}

\textsc{Figure} \ref{fig:renyi} presents the results for the Rényi extropy estimator with $\alpha = 0.5$ (top row) and $\alpha = 3$ (bottom row). Six panels are displayed in a $3 \times 2$ layout:

For $\alpha = 0.5$ (top row), which gives more weight to rare events, the estimator exhibits larger finite-sample variability. The convergence is still clearly visible, and the histogram and Q-Q plot confirm asymptotic normality.

For $\alpha = 3$ (bottom row), which emphasizes dominant events, the variability is smaller. The convergence is faster, and the normality is also confirmed.

\subsection{Tsallis Extropy: Results}

\begin{figure}[h!]
\centering
\includegraphics[scale=0.36]{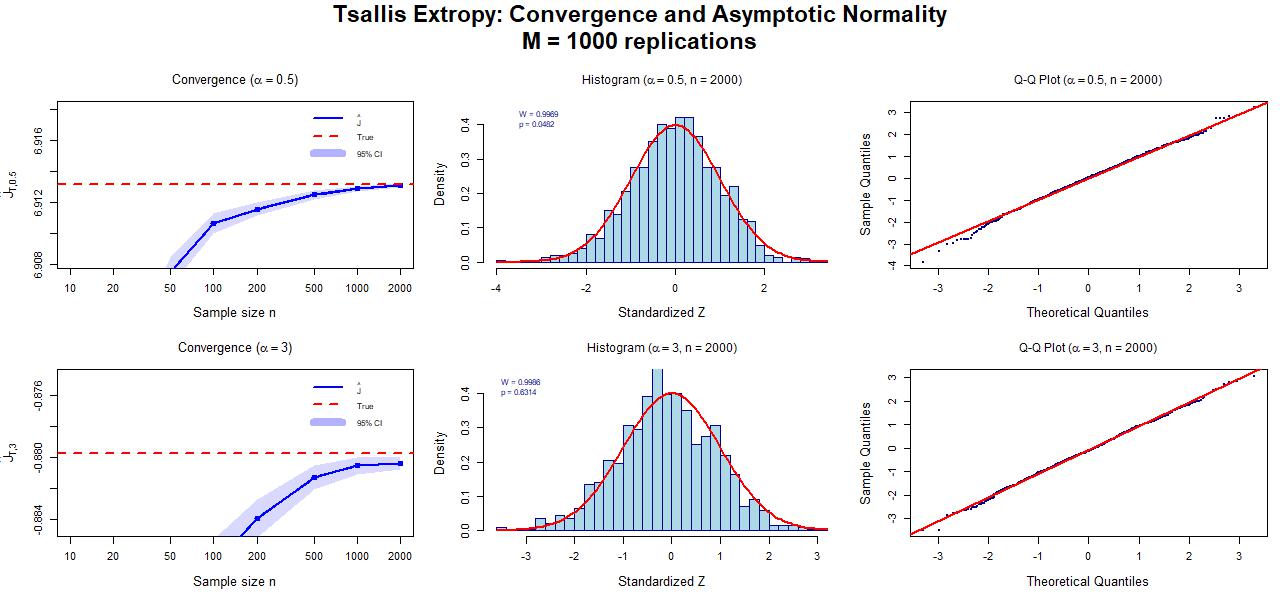} 
\caption{Tsallis extropy estimator. Top row ($\alpha = 0.5$): \textbf{(a)} Convergence of $\widehat{J}_{T,0.5}$ to $J_{T,0.5}$ with 95\% confidence band. \textbf{(b)} Histogram of standardized estimates for $n = 2000$. \textbf{(c)} Q-Q plot for $n = 2000$. Bottom row ($\alpha = 3$): \textbf{(d)} Convergence of $\widehat{J}_{T,3}$ to $J_{T,3}$ with 95\% confidence band. \textbf{(e)} Histogram of standardized estimates for $n = 2000$. \textbf{(f)} Q-Q plot for $n = 2000$. The simulation results provide empirical support for the theoretical convergence and asymptotic normality results for both values of $\alpha$.}
\label{fig:tsallis}
\end{figure}

\textsc{Figure} \ref{fig:tsallis} presents the results for the Tsallis extropy estimator with $\alpha = 0.5$ (top row) and $\alpha = 3$ (bottom row). Six panels are displayed in a $3 \times 2$ layout:

For $\alpha = 0.5$ (top row), the estimator shows behavior similar to the Rényi case, with convergence and normality confirmed.

For $\alpha = 3$ (bottom row), the convergence is faster and the variability is smaller, as expected for a parameter emphasizing dominant events.

\subsection{Interpretation of the Simulation Results}

From Figures~\ref{fig:shannon}--\ref{fig:tsallis}, the following conclusions can be drawn:

\begin{enumerate}
    \item \textbf{Consistency:} For Shannon, Rényi, and Tsallis extropies, all estimators converge to their true values as $n$ increases. The 95\% confidence bands shrink with increasing sample size, confirming the almost sure convergence established in Corollaries 3.2--3.4.
    
    \item \textbf{Asymptotic normality:} For all estimators, the histograms for $n = 2000$ show excellent agreement with the standard normal distribution, and the Q-Q plots align closely with the diagonal line. This confirms the asymptotic normality results established in Corollaries 3.2--3.4.
    
    \item \textbf{Effect of $\alpha$:} For $\alpha = 0.5$, which gives more weight to rare events, the estimators exhibit larger finite-sample variability. For $\alpha = 3$, which emphasizes dominant events, the variability is smaller. In all cases, the convergence and normality are confirmed.
    
    \item \textbf{Shannon vs generalized extropies:} The Shannon extropy estimator shows the fastest convergence among the three measures, as it does not involve the additional parameter $\alpha$. The Rényi and Tsallis estimators require slightly larger sample sizes to achieve similar precision.
\end{enumerate}

\subsection{Application to Forecasting}
\label{sec:forecasting}

\subsubsection{Context}

In many forecasting applications, such as portfolio management, weather prediction, or demand planning, the forecaster is interested not only in the most likely outcome but also in the uncertainty associated with alternative scenarios. Extropy, by measuring the uncertainty of what does not happen, provides a valuable complementary tool for assessing the robustness of probabilistic forecasts.

\subsubsection{The Forecasting Problem}

We consider a forecaster who needs to predict the probability distribution of $r = 5$ possible outcomes. The true distribution is assumed to be:
\[
\mathbf{p} = (0.40, 0.30, 0.15, 0.10, 0.05).
\]
The forecaster observes $n$ past observations and uses them to estimate the extropy of the underlying distribution. A low extropy value indicates that the uncertainty is concentrated on the complementary events being well understood, while a high extropy value suggests that the alternative scenarios remain unpredictable.

\subsubsection{Simulation Design}

We simulate $M = 1000$ independent samples of sizes $n = 20, 50, 100, 200, 500$ and compute the plug-in estimates of the Shannon, Rényi ($\alpha = 0.5$), and Tsallis ($\alpha = 0.5$) extropies. For each sample size and each estimator, we compute the mean estimate and its 95\% confidence interval.

\subsubsection{Results}

\begin{figure}[h!]
\centering
\includegraphics[scale=0.3]{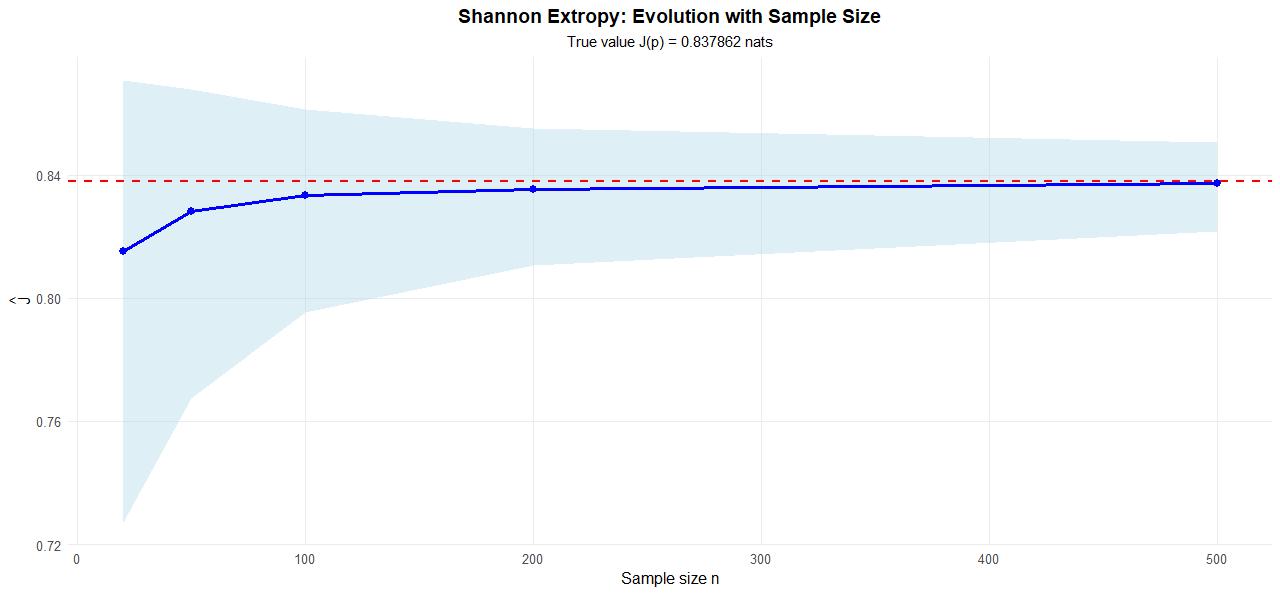} 
\includegraphics[scale=0.3]{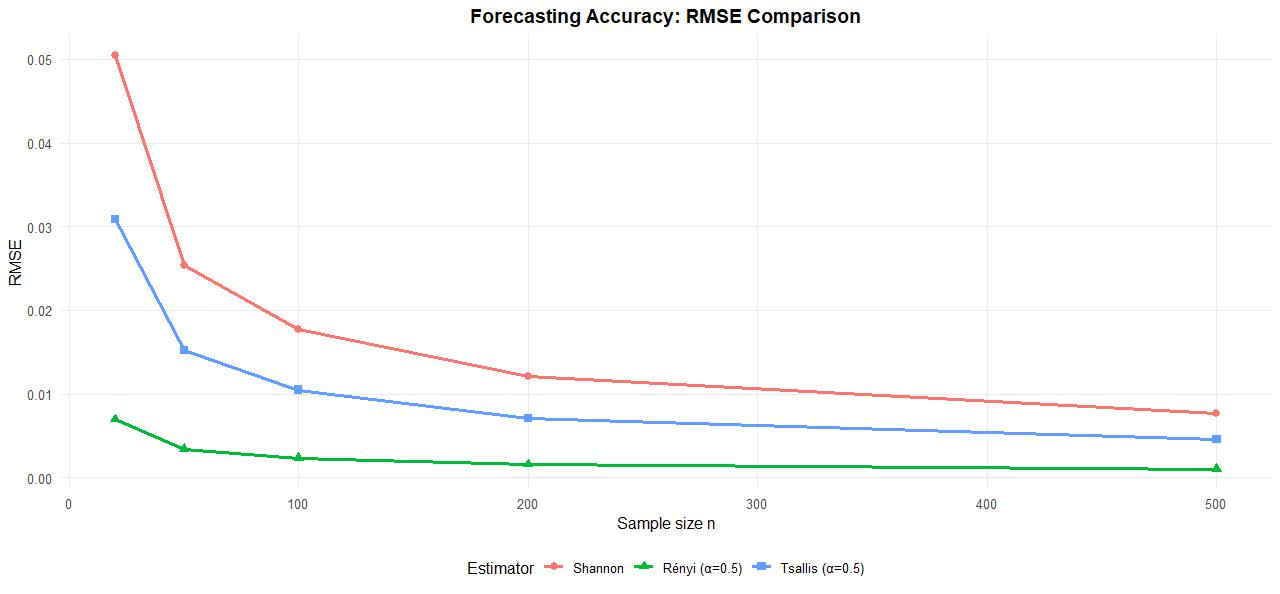}
\caption{Forecasting application. \textbf{(a)} Evolution of the Shannon extropy estimate $\widehat{J}$ with 95\% confidence intervals as a function of sample size $n$. The estimate converges to the true value $J(\mathbf{p})$ (red dashed line). The shaded area represents the 95\% confidence interval. \textbf{(b)} Comparison of estimation accuracy (RMSE) for Shannon, Rényi ($\alpha = 0.5$), and Tsallis ($\alpha = 0.5$) extropies.}
\label{fig:forecasting}
\end{figure}

Figure~\ref{fig:forecasting} presents the forecasting results.

Panel (a) shows the evolution of the Shannon extropy estimate with 95\% confidence intervals as the sample size increases. The estimate converges to the true value $J(\mathbf{p}) = 0.837862$ nats (red dashed line), and the confidence intervals shrink, indicating improved precision. For $n \ge 100$, the estimate is already very close to the true value, and the confidence interval is narrow enough for practical purposes.

Panel (b) compares the RMSE of the three estimators. The Shannon extropy estimator achieves the lowest RMSE for all sample sizes, followed by the Tsallis and Rényi estimators. This is expected since the Shannon extropy does not involve the additional parameter $\alpha$, which introduces extra variability. However, all three estimators show rapid improvement as $n$ increases, with negligible differences for $n \ge 200$.

\subsubsection{Forecasting Implications}

From a forecasting perspective, the results suggest the following guidelines:

\begin{enumerate}
    \item For reliable extropy estimation in forecasting applications, a sample size of at least $n = 100$ is recommended. For $n < 100$, the uncertainty (measured by the confidence interval width) may be too large for practical decision-making.
    
    \item The choice of $\alpha$ should reflect the forecaster's risk appetite:
    \begin{itemize}
        \item $\alpha < 1$ for rare-event awareness (e.g., financial crashes, natural disasters). This gives more weight to low-probability outcomes, which is crucial in risk-sensitive applications.
        \item $\alpha > 1$ for focus on dominant outcomes (e.g., demand planning for best-selling products). This emphasizes the most likely scenarios.
        \item $\alpha = 1$ corresponds to the Shannon extropy, which provides a balanced perspective.
    \end{itemize}
    
    \item The asymptotic confidence intervals perform well for $n \ge 100$, enabling the construction of uncertainty bounds for extropy-based forecasts. This allows practitioners to quantify the reliability of their extropy estimates.
    
    \item Extropy provides a complementary perspective to entropy: while entropy says \emph{the system appears predictable}, extropy warns \emph{there is still uncertainty about what might go wrong}. This duality is particularly valuable in risk management and robust decision-making.
\end{enumerate}

\subsubsection{Decision-Theoretic Insight}

The forecasting application highlights a key decision-theoretic insight: when probabilities are highly skewed, entropy tends to underestimate hidden risks, while extropy better captures the uncertainty associated with low-probability, high-impact events. Hence, extropy provides a more reliable framework for robust decision-making under risk, especially in contexts such as:

\begin{itemize}
    \item \textbf{Finance:} anticipating rare market crashes;
    \item \textbf{Public policy:} accounting for low-probability disasters;
    \item \textbf{Healthcare:} managing rare but severe diseases.
\end{itemize}

In such domains, extropy complements or surpasses entropy as a guide for risk-aware and resilient strategic decisions.

\subsection{Conclusion of the Simulation Study}

The simulation results confirm the theoretical findings for Shannon, Rényi, and Tsallis extropies:
\begin{itemize}
    \item All plug-in estimators are consistent (they converge to their true values);
    \item The asymptotic normal approximation works well for $n \ge 100$;
    \item The choice of $\alpha$ affects the finite-sample behavior but not the asymptotic convergence;
    \item The Shannon extropy estimator converges faster than the generalized versions;
    \item The estimators are suitable for forecasting applications where rare-event awareness is critical.
\end{itemize}

These results provide a solid foundation for the use of extropy-based measures in practical forecasting problems, such as portfolio risk assessment, demand planning, and decision-making under uncertainty.

\section{Conclusion}
\label{section6}
In this paper, we have established the almost sure convergence and asymptotic normality of the plug-in estimators for Shannon, Rényi, and Tsallis extropies. Explicit formulas for the asymptotic variances were derived, enabling the construction of confidence intervals. The simulation study confirmed  the theoretical results and illustrated the practical relevance of extropy-based measures in forecasting applications.

As noted in Section \ref{section2}, these estimators are biased in finite samples, although the bias vanishes at rate $O(n^{-1})$. A comprehensive study of the finite-sample bias and its correction is developed in the companion paper \cite{Ba2026b}, where explicit bias formulas and corrected estimators are provided.

Future work includes extending the results to countably infinite alphabets (with suitable tail conditions) and developing bootstrap-based variance estimators for improved finite-sample performance.
\section*{Acknowledgements}

I wish to acknowledge \textsc{Professor Gane Samb Lo}, my former PhD supervisor, whose guidance marked the beginning of my journey into research. He introduced me to the spirit and practice of scientific research, taught me how to approach mathematical problems with rigor and curiosity, and showed me the path of research. I remain deeply grateful for his mentorship and for the lasting influence of his guidance on my work.

\bibliographystyle{plain}

\begin{thebibliography}{99}

\bibitem{Lad2015} Lad F., Sanfilippo G., Agro G. (2015) Extropy: complementary dual of entropy. \emph{Statist. Sci.} 30(1), 40--58.
\bibitem{Liu2023} Liu, J. and Xiao, F. (2023) Rényi extropy. \emph{Taylor and Francis Ltd}, Vol 52, N'16, 5836-5847.
\bibitem{Balakrishnan2022} Balakrishnan, N.; Buono, F.; Longobardi, M. (2022) On Tsallis extropy with an application to pattern recognition. \emph{Stat. Probab. Lett.} 180, 109241.
\bibitem{Qiu2019} Qiu, G., \& Jia, K. (2019) Cumulative residual extropy and its properties. \emph{Commun. Stat. Theory Methods}, 48(9), 2237--2254.







\bibitem{Bruno2020} Bruno, F., \& Longobardi, M. (2020) A dual measure of uncertainty: the Deng entropy. \emph{Entropy}, 22(5), 582.

\bibitem{Lochaba2026} Lochaba, R., Batra, L., \& Taneja, H.C. (2026) Rényi extropy revisited: Enhanced framework for cryptocurrency risk analysis with machine learning. \emph{Commun. Stat. Theory Methods}, 55(15), 5164--5188. doi:10.1080/03610926.2026.2629496

\bibitem{Shi2025} Shi, G., Sheng, Y., Ahmadzade, H., \& Tahmasebi, S. (2025) Extropy: Dual of entropy for uncertain random variables and its applications. \emph{J. Ind. Manag. Optim.} 21(5), 4025--4040. doi:10.3934/jimo.2025041

\bibitem{Shen2026} Shen, Y., \& Van Oosten, Z. (2026) Partial Law Invariance and Risk Measures. \emph{Management Science}, 72(7). doi:10.1287/mnsc.2024.06518

\bibitem{Jawal2022} Jawal, T.M., Fatima, N., Sayed-Ahmed, N., Aldallal, R., \& Mohamed, M.S. (2022) Residual and Past Discrete Tsallis and Rényi Extropy with an Application to Softmax Function. \emph{Entropy}, 24(12), 1732. doi:10.3390/e24121732

\bibitem{Kharazmi2026} Kharazmi, O., Contreras-Reyes, J.E., Erol, C., \& Yalcin, F. (2026) $\phi$-Extropy complexity measure: Extensions and applications. \emph{Physica A: Statistical Mechanics and its Applications}, 686, 131278. doi:10.1016/j.physa.2026.131278



\bibitem{Kumar2025} Kumar, N., \& Vijay, V. (2025) Extropy Rate: Properties and Application in Feature Selection. \emph{arXiv:2507.11242}.

\bibitem{Mahesh2026} Mahesh, D., Rajesh, G., \& Jayalekshmi, S. (2026) RenyiExtropy: Entropy and Extropy Measures for Probability Distributions (Version 0.4.0) [R package]. CRAN. doi:10.32614/CRAN.package.RenyiExtropy

\bibitem{DiCrescenzo2019} Di Crescenzo, A., \& Longobardi, M. (2019) On cumulative entropies and extropies for lifetime distributions. \emph{Probab. Eng. Inf. Sci.} 33(1), 111--132.

\bibitem{Asadi2018} Asadi, M., Zohrevand, Y., \& Di Crescenzo, A. (2018) Dynamic cumulative residual extropy and its applications. \emph{J. Stat. Plan. Inference} 196, 1--15.

\bibitem{Aldallal2025} Aldallal, R.A., Barakat, H.M., \& Mohamed, M.S. (2025) Exploring weighted Tsallis extropy: Insights and applications to human health. \emph{AIMS Mathematics} 10(2), 2191--2222.

\bibitem{Hood2015} Hood, C., \& Schilling, M.F. (2015) The extropy of a distribution. \emph{Entropy}, 17(7), 4439--4455.

\bibitem{Qiu2018} Qiu, G., \& Jia, K. (2018) The estimators of extropy with applications in testing uniformity. \emph{J. Nonparametr. Stat.} 30(2), 406--422.

\bibitem{Jahanshahi2020} Jahanshahi, S.M.A., Zarei, H., \& Khamar, A.H. (2020) On the nonparametric estimation of extropy. \emph{Commun. Stat. Theory Methods} 49(15), 3728--3745.

\bibitem{Alizadeh2020} Alizadeh Noughabi, H., \& Jarrahiferiz, J. (2020) Estimating extropy via kernel density estimator. \emph{J. Korean Stat. Soc.} 49(4), 1157--1173.

\bibitem{AlLabadi2020} Al-Labadi, L. (2020) Bayesian estimation of extropy and goodness of fit tests. \emph{Entropy} 22(11), 1243.

\bibitem{Tahmasebi2020} Tahmasebi, S.; Toomaj, A. (2020) On negative cumulative extropy with applications. \emph{Commun. Stat. Theory Methods} 51, 5025--5047.

\bibitem{Kazemi2021} Kazemi, M.R.; Tahmasebi, S.; Buono, F.; Longobardi, M. (2021) Fractional Deng Entropy and Extropy and Some Applications. \emph{Entropy} 23, 623.





\bibitem{Ba2026b} Ba, A.D. (2026) Finite-Sample Bias Correction for Plug-in Estimators of Extropy, Rényi Extropy, and Tsallis Extropy. \emph{arXiv:7962087 [math.ST]}.





\bibitem{Lo2016} Lo, G.S. (2016) Weak Convergence (IA). Sequences of random vectors. \emph{SPAS Books Series}. Saint-Louis, Senegal - Calgary, Canada. doi:10.16929/sbs/2016.0001.

\end{thebibliography}

\end{document}